\documentclass[preprint,3p,twocolumn,nofootinbib]{elsarticle}
\usepackage{epsfig}
\usepackage{bm}
\usepackage{latexsym}
\usepackage{natbib}
\usepackage{url}
\usepackage{dcolumn}
\usepackage{color}
\usepackage[usenames,dvipsnames]{xcolor}
\usepackage{amsfonts,amssymb,amsmath}
\usepackage{graphicx,epsfig}
\usepackage{psfrag}
\usepackage{subfigure}
\usepackage{comment}
\usepackage[utf8]{inputenc}
\usepackage{amsmath}
\usepackage{hyperref}
\hypersetup{colorlinks=true}

\def\be{\begin{equation}}
\def\ee{\end{equation}}
\def\gev{{\rm \,Ge\kern-0.125em V}}

\begin{document}
 
\title{Distance, de Sitter and Trans-Planckian Censorship conjectures:\\
the status quo of Warm Inflation}

\author[1]{Suratna Das}%
 \ead{suratna@iitk.ac.in}
\address[1]{Department of Physics, Indian Institute of Technology, Kanpur 208016, India}

\date{\today}

\begin{abstract}
In this short note, we pointed out that Warm inflationary scenario remains to be favoured over its cold counterpart by the recently proposed conjectures, which aim to overrule de Sitter like constructions in String Landscapes. On the other hand, the canonical cold (single-field slow-roll) inflationary models, which were in tension with the previously proposed de Sitter conjecture, have now become even more unlikely to realise in String Landscapes as these scenarios fail to cope up with both the conjectures, de Sitter and trans-Planckian censorship, at one go. 
\end{abstract}

\maketitle

Over the past one year, the {\it Swampland Conjectures}, proposed in \cite{Obied:2018sgi, Garg:2018reu, Ooguri:2018wrx}, have been much debated in the literature \cite{Kachru:2018aqn, Akrami:2018ylq}, and it was rightfully so as these conjectures tend to overrule any construction of stable or meta stable de Sitter vaccum in the String Landscapes, yielding it difficult to realise inflationary as well as dark energy dominated universes, two cornerstones of the standard model of cosmology, in quantum field theories compatible with quantum gravity theories, such as String theory \cite{Akrami:2018ylq, Agrawal:2018own, Agrawal:2018rcg, Raveri:2018ddi}. A new conjecture, the {\it Trans-Planckian Censorship} \cite{Bedroya:2019snp}, has recently been proposed along the same lines, which tends to constrain the form of the scalar potentials, like its preceder.  In this conjecture, the well-known {\it trans-Planckain problem} of inflationary cosmology \cite{Martin:2000xs, Easther:2001fi}, which raises the concern of crossing the Hubble radius by sub-Planckian perturbations during inflation responsible for seeding the structures of the universe at later stages, has been aimed to be avoided in the String Landscapes. The `immidiate' cosmological consequences of this newly proposed conjecture has been discussed in \cite{Bedroya:2019tba, Tenkanen2019, Cai:2019hge}.

A quick glance at these recently proposed conjectures and the bounds set by them would help us gauge the present scenario:
\begin{itemize}
\item {\bf The distance conjecture:} This criteria limits the field range traversed by a scalar field in the field space as \cite{Obied:2018sgi}
\begin{eqnarray}
\frac{\Delta\phi}{M_{\rm Pl}}\lesssim d,
\end{eqnarray}
where $d\sim{\mathcal O}(1)$.\footnote{However, it was shown in \cite{Scalisi:2018eaz} that taking the Hubble scale during inflation smaller than the cutoff scale of quantum gravity the bound translates into $\Delta\phi/M_{\rm Pl}\lesssim -\log{r}$, $r$ being the tensor-to-scalar ratio.} This criteria favours small field inflation over the large field ones. 

\item {\bf The (refined) de Sitter conjecture:} This condition puts bound on the form of the scalar potential $V(\phi_i)$ as \cite{Garg:2018reu, Ooguri:2018wrx}
\begin{eqnarray}
M_{\rm Pl}\frac{|V_\phi|}{V}>c\quad{\rm or}\quad M_{\rm Pl}^2\frac{{\rm min}(V_{\phi_i\phi_j})}{V}\leq -c',
\end{eqnarray}
where $V_\phi\equiv \partial V(\phi)/\partial\phi$, and $c,c'\sim\mathcal{O}(1)$. 

For canonical single field inflation (with canonical kinetic term and Bunch-Davies vacuum), which would be the case under consideration in this note,\footnote{Other extensions of the canonical picture, such as non-Bunch-Davies vacuum \cite{Ashoorioon:2018sqb} or non-canonical kinetic term \cite{Das:2018hqy, Mizuno:2019pcm}, are beyond the scope of this discussion} the above bounds on scalar field potentials readily translate into bounds on the slow-roll parameters as 
\begin{eqnarray}
\epsilon_\phi&\equiv& \frac{M_{\rm Pl}^2}{2}\left(\frac{V_\phi}{V}\right)^2>\frac{c^2}{2},\nonumber\\
 &&\quad {\rm or}\quad \nonumber\\
 \eta_\phi&\equiv& M_{\rm Pl}^2\frac{V_{\phi\phi}}{V}\leq -c'.
\end{eqnarray}
These bounds contravene either of the two slow-roll conditions of inflationary dynamics and are also in conflict with the present observations \cite{Das:2018hqy}. On the one hand, $\epsilon_\phi>\mathcal{O}(1)$ (in which case $\eta_\phi$ can be less than unity) yields too large tensor-to-scalar ratio $r=16\epsilon_\phi$ \cite{Agrawal:2018own} to be in accordance with the present bound $r<0.065$  \cite{Akrami:2018odb}, and on the other hand, $|\eta_\phi|>\mathcal{O}(1)$ (in which case $\epsilon_\phi$ can be less than unity) would yield a tilted scalar spectrum $n_s-1=2\eta_\phi-6\epsilon_\phi\sim 2\eta_\phi$ \cite{Fukuda:2018haz}, ruled out by the present observation $n_s\sim 0.9649\pm0.0042$  \cite{Akrami:2018odb}. Thus the (refined) de Sitter conjecture directly challenges construction of single-field slow-roll dynamics in String Landscapes. 

\item {\bf The trans-Planckian censorship conjecture:} Aiming that no length scale smaller than the Planck length $l_{\rm Pl}$ should leave the Hubble horizon during a de Sitter evolution, it was straightforwardly conjectured that \cite{Bedroya:2019snp}
\begin{eqnarray}
\left(\frac{a_f}{a_i}\right)l_{\rm Pl}<\frac{1}{H_f},
\end{eqnarray} 
where $a_f$ and $a_i$ are the scale factors at the beginning and end of the evolution and $H_f^{-1}$ is the Hubble radius at the end of such evolution. The above equation can be more conveniently written as 
\begin{eqnarray}
e^N\sim \frac{a_f}{a_i}<\frac{M_{\rm Pl}}{H_f},
\label{trans-planck}
\end{eqnarray} 
where $N$ is the number of e-foldings. This bound has then been further massaged to put bounds on the scalar potentials  in field theories residing  in the String Landscapes \cite{Bedroya:2019snp}. 

For large field-range scenarios, this leads to the bound 
\begin{eqnarray}
M_{\rm Pl}\frac{|V_\phi|}{V}\geq \frac{2}{\sqrt6},
\end{eqnarray}
which is in tune with the first of the two conditions of the refined de Sitter conjecture. Hence for large-field range scenarios, the quantum field theory resides in String Landscapes, i.e. satisfying both de Sitter and trans-Plackian censorship conjectures, if only the first of the two de Sitter conjecture conditions is maintained, i.e. $\epsilon_\phi>1$ and $\eta_\phi<1$. However, such scenarios are prone to be in tension with the distance conjecture, and violate the observational constraint on $r=16\epsilon_\phi<0.065$. 

For short field-range scenarios the trans-Planckian censorship conjecture imposes a milder constraint as 
\begin{eqnarray}
M_{\rm Pl}\frac{|V_\phi|_{\rm max}}{V_{\rm max}}>\frac{\phi_f-\phi_i}{24}\ln\left(\sqrt{\frac{3}{V_f}}\right)^{-2},
\end{eqnarray}
where $|V_\phi|_{\rm max}$ is the maximum of $V_{\phi}$ over the field range $[\phi_i,\phi_f]$. Thus, the short field-range cases, which also obey the distance conjecture, can satisfy either of the refined de Sitter conjectures and reside in the field theories within String Landscapes. However, as has been pointed out previously, the single-field slow-roll inflationary scenario is in tension with both the refined de Sitter conjecture conditions, and hence even the short-range scenarios doesn't serve the purpose any better. 
\end{itemize}
In a nutshell, neither the long-ranged nor the short-ranged single-field slow-roll scenarios within the cold inflation paradigm satisfy all the Swampland conjectures, distance, de Sitter and the trans-Planckian censorship conjecture, at one go, and hence cannot be realised in String Landscapes.   

It was first observed in \cite{Das:2018hqy} and was later illustrated in \cite{Das:2018rpg, Motaharfar:2018zyb} that Warm inflationary scenario \cite{Berera:1995ie} is capable of evading the distance as well as the de Sitter conjectures in contrast to the existing cold single-field models. It was due to the dissipative dynamics of the inflaton field during Warm inflation, 
\begin{eqnarray}
\ddot\phi+3H\dot\phi+V_{\phi}=\Gamma\dot\phi, 
\end{eqnarray}
that allows the slow-roll parameters to retain values larger than unity $(\epsilon_\phi, \eta_\phi\ll1+Q$, where $Q\equiv \Gamma/3H$, $\Gamma$ being the decay rate of the inflaton field into a constant radiation bath) during inflation, helping the scenario avoid the stringent constraints imposed by the (refined) de Sitter conjectures. However, it was noted both in \cite{Das:2018rpg} and in \cite{Motaharfar:2018zyb} that the Warm inflationary scenario evades these conjectures more efficiently in the strong dissipative regime, when $Q\gg1$. 

It was a tasking job to realise Warm inflation in a concrete particle physics model \cite{Yokoyama:1998ju}, and the stumbling block has  recently been overcome in Warm Little Inflaton model \cite{Bastero-Gil:2016qru} where the inflaton field is a pseudo Nambu-Goldstone boson coupled to a pair of light fermionic fields. However, this scenario, with $\Gamma\propto T$ where $T$ is the temperature of radiation bath during inflation, is compatible with the observations only in weak dissipative regime $(Q\ll1)$. It is due to the fact that $d\Gamma/dT>0$ results in a substantial enhancement of inflaton fluctuations in the strong dissipative regime with $Q>1$ which then yields a blue-tilted spectrum, ruled out by observations. The inflaton perturbations would be damped in a strong dissipative regime only if $d\Gamma/dT<0$ is achieved. A Warm inflationary model \cite{Bastero-Gil:2019gao}, much on the lines of the Warm Little Inflaton model, has been very recently devised where the pseudo Nambu-Goldstone boson inflaton couples also to a pair of bosonic fields, yielding  $\Gamma\propto T^{-1}$, and hence allowing for $Q$ as large as $100$ for a quadratic inflaton potential. 
Thus, it is obvious that such Warm inflation models are in favour of realising inflationary dynamics in String Landscapes. It is also worthwhile to note that in strong dissipative regime the much-debated $\eta$-problem \cite{Baumann:2014nda} of String Theory is also greatly alleviated \cite{Bastero-Gil:2019gao}. 

The newly devised trans-Planckian censorship conjecture has thrown in new challenges to the inflationary dynamics by not only constraining the scalar potentials, but also severely bringing down the scale of inflation, as has been pointed out in \cite{Bedroya:2019tba}. It was shown that the trans-Planckian censorship condition stated in Eq.~(\ref{trans-planck}) can be translated into a bound on the energy density (or energy scale) during inflation as 
\begin{eqnarray}
V^{1/4}<6\times 10^8\,{\rm GeV}\sim 3\times 10^{-10} M_{\rm Pl}.
\label{v-bound}
\end{eqnarray}
This bound is independent of the nature of the dynamics that drives inflation. In generic cold inflation model, driven by a single field, the scalar and tensor power spectra turn out to be 
\begin{eqnarray}
\mathcal{P}_{\mathcal R}&=&\frac{1}{8\pi^2}\left(\frac{H^2}{\epsilon_\phi M_{\rm Pl}^2}\right)\left(\frac{k}{k_*}\right)^{n_s-1}\nonumber\\
&\equiv& A_s(k_*)\left(\frac{k}{k_*}\right)^{n_s-1},\nonumber\\
\mathcal{P}_T&=&\frac{2}{\pi^2}\left(\frac{H^2}{M_{\rm Pl}^2}\right)\left(\frac{k}{k_*}\right)^{n_T}\nonumber\\
&\equiv& A_T(k_*)\left(\frac{k}{k_*}\right)^{n_T},
\end{eqnarray}
respectively, where $k_*=0.002$ Mpc$^{-1}$ is the pivot scale where the amplitudes of these spectra are measured. Using the Friedmann equation during inflation, $3H^2=V/M_{\rm Pl}^2$, the observation of scalar amplitude $A_s=2.215\times 10^{-9}$ \cite{Akrami:2018odb} and the bound on $V(\phi)$  coming from the trans-Planckian censorship conjecture, as stated in Eq.~(\ref{v-bound}), one  constrains the slow-roll parameter $\epsilon_\phi$ severely as \cite{Bedroya:2019tba}
\begin{eqnarray}
\epsilon_\phi\lesssim 10^{-32}.
\label{epsilon}
\end{eqnarray}
Though, the slow-roll parameter $\epsilon_\phi$ is not directly observable, the observable quantity $r$, the tensor-to-scalar-ratio which is the ratio of the amplitudes of the tensor and the scalar powers, is proportional to this slow-roll parameter ($r=16\epsilon_\phi$). Thus a stringent bound on $\epsilon_\phi$ coming from the trans-Planckian censorship conjecture puts an upper bound on this observable as \cite{Bedroya:2019tba}
\begin{eqnarray}
r\lesssim 10^{-31}.
\end{eqnarray}
Hence, any detection of $r$ by the near future observations like BICEP3 \cite{Wu:2016hul} and LiteBIRD \cite{Matsumura:2013aja}, which will seek for tensor-to-scalar ratio within the range $r\sim 0.001\,-\,0.01$, would lead to a tension with the newly proposed trans-Planckian censorship conjecture. 

We will add another remark which has not been mentioned previously in \cite{Bedroya:2019tba, Tenkanen2019}. The bound on $\epsilon_\phi$, as stated in Eq.~(\ref{epsilon}),  points out that 
the first of the two conditions of the refined de Sitter conjectures should be violated in single field inflation, if trans-Planckian censorship conjecture is to be respected. Moreover, in such a case, when $\epsilon_\phi\ll1$, the observation of scalar spectral index $n_s-1\sim 2\eta_\phi$ would demand $|\eta_\phi|\ll1$, violating the second option of the refined de Sitter conjecture as well. Therefore, it is important to note that canonical cold single-field models fail to evade both the conjectures, de Sitter as well as trans-Planckian censorship, at one go and hence cannot be realised in any field theory residing in String Landscapes. 

In Warm inflationary model, the scalar power in the strong dissipative regime turns out to be \cite{Bastero-Gil:2019gao}
\begin{eqnarray}
{\mathcal P}_{\mathcal R}\simeq \frac{\sqrt{3\pi}}{24\pi^2}\frac{V(\phi_*)}{\epsilon_\phi M_{\rm Pl}^4}\left(\frac{T_*}{H_*}\right)Q_*^{5/2}\left(\frac{Q_*}{Q_c}\right)^{\beta_c},
\end{eqnarray}
where $1\lesssim Q_c\lesssim 10$, and $\beta_c\sim -1.6$ for $\Gamma\propto T^{-1}$, with a scalar spectral index  \cite{Ramos:2013nsa}
\begin{eqnarray}
n_s-1\sim\frac{3}{4Q}\left(-3\epsilon_\phi+2\eta_\phi-3\beta_\phi\right),
\end{eqnarray}
where $\beta_\phi\equiv M_{\rm Pl}^2 (\Gamma_\phi V_\phi)/(\Gamma V)$ is another slow-roll parameter in Warm inflationary dynamics. Besides, as the tensor perturbations are not affected by the dissipative dynamics of the Warm inflation, the tensor spectrum retains the same form as that of in cold inflation, yielding the tensor-to-scalar ratio in Warm inflation as 
\begin{eqnarray}
r\simeq \frac{16\epsilon_\phi}{\sqrt{3\pi}Q_*^{5/2}}\left(\frac{H_*}{T_*}\right)\left(\frac{Q_*}{Q_c}\right)^{1.6}.
\end{eqnarray}

The bound on the potential energy $V$ coming from the trans-Planckian censorship conjecture (Eq.~(\ref{v-bound})), which is independent of the cold or warm dynamics of inflation, will drive the slow-roll parameter $\epsilon_\phi$ to very small values even in the case of Warm inflation. In such a case the scalar spectral tilt would become $n_s-1\sim 3\eta_\phi/2Q$ (ignoring $\beta_\phi$ by assuming $\Gamma \neq \Gamma(\phi)$). Thus, one requires $Q\sim 20$ with $|\eta_\phi|\sim \mathcal{O}(1)$ to meet the observed value of the spectral index $n_s\sim 0.9649$ \cite{Akrami:2018odb}. Therefore, with $Q\sim 20$, $T_*\sim10^3 H_*$ and $Q_c\sim10$ the bound on the potential energy from trans-Planckian censorship would constrain $\epsilon_\phi$ as well as $r$ as 
\begin{eqnarray}
\epsilon_\phi\lesssim 10^{-26},\quad {\rm and}\quad r\lesssim 10^{-31}.
\end{eqnarray}
Therefore, the warm inflationary scenario, unlike its cold counterpart, can be in accordance with the trans-Planckian censorship conjecture as well as with the refined de Sitter conjecture for potentials yielding $\epsilon_\phi\ll1$ and $\eta_\phi\sim\mathcal{O}(1)$.
In addition, we note that the field traversed during Warm inflation (in strong dissipative regime) can be written as \cite{Das:2018rpg}
\begin{eqnarray}
\frac{\Delta\phi}{M_{\rm Pl}}\sim \sqrt{\frac r8\left(\frac TH\right)Q^{1/2}}\Delta N,
\end{eqnarray}
which yields 
\begin{eqnarray}
\Delta\phi\sim10^{-12}M_{\rm Pl},
\end{eqnarray}
indicating small-field dynamics which are in tune with the distance conjecture. Also, in this regime the trans-Planckian censorship conjecture puts milder bounds on the form of the inflaton potentials, as mentioned in the beginning. 

It is well known that the small-field models suffer from initial condition problem as they are not local attractors in the initial condition space and fine-tuning of the kinetic energy of the inflaton field is called for to enter a slow-rolling regime (for a recent review see \cite{Chowdhury:2019otk}). This issue has also been pointed out in \cite{Bedroya:2019tba} as a consequence of the constraints put by the trans-Planckian censorship conjecture. Here we note that simple manipulations yield the fine-tuning of the inflaton kinetic energy $\dot\phi_{SR}$ to enter the slow-roll dynamics to be  
\begin{eqnarray}
\frac{\dot\phi_{SR}}{\dot\phi_i}\sim \frac{\epsilon_\phi^{1/2}}{(1+Q)^{1/2}}\sim 10^{-13},
\end{eqnarray}
demanding that the initial kinetic energy $\phi_i\sim V$. But the fluctuation-dissipation dynamics \cite{Bastero-Gil:2016mrl} which prevails in the pre-inflationary era of Warm inflation is much more rich and thus a thorough analysis of small-field dynamics in Warm inflation is called for to concretely gauge the situation. 

{\it In conclusion:} A quantum field theory capable of evading all the three conjectures, distance, de Sitter and trans-Planckian censorship, would reside in a String Landscape and would be compatible with quantum gravity theories. We pointed out here, that the canonical cold inflationary models, which were previously in tension with the refined de Sitter conjecture \cite{Das:2018hqy}, have now become incompatible with the two conjectures, de Sitter and trans-Planckian censorship, taken together and thus are unlikely to qualify to reside in String Landscapes any longer. However, the Warm inflationary scenario in the strong dissipative regime, which was quite in tune with the de Sitter conjectures previously, has now shown to be compatible with all the three above mentioned conjectures at one go, and thus is more likely to be realised in String Landscapes. Even though, one still needs to figure out the Warm inflationary potential which would yield a negligible $\epsilon_\phi$ and $|\eta_\phi|\sim \mathcal{O}(1)$ to make the scenario acceptable. We defer this analysis for a future endeavour. Above all, it is well known that small-field models, which are now preferred by both distance and trans-Planckian censorship conjectures, suffer from initial conditions problems and a severe fine tuning of the initial kinetic energy is required in such cases to onset inflation. The pre-inflationary dynamics of Warm inflation is much richer than the generic cold inflation and a thorough analysis of the initial conditions problem of small field models in Warm inflation is now called for. 
\\

{\it Acknowledgements}:
 The work of S.D. is supported by Department of Science and Technology, 
 Government of India under the Grant Agreement number IFA13-PH-77 (INSPIRE Faculty Award).

\label{Bibliography}
\bibliography{conjectures}

\begin{thebibliography}{32}
\expandafter\ifx\csname natexlab\endcsname\relax\def\natexlab#1{#1}\fi
\providecommand{\url}[1]{\texttt{#1}}
\providecommand{\href}[2]{#2}
\providecommand{\path}[1]{#1}
\providecommand{\DOIprefix}{doi:}
\providecommand{\ArXivprefix}{arXiv:}
\providecommand{\URLprefix}{URL: }
\providecommand{\Pubmedprefix}{pmid:}
\providecommand{\doi}[1]{\href{http://dx.doi.org/#1}{\path{#1}}}
\providecommand{\Pubmed}[1]{\href{pmid:#1}{\path{#1}}}
\providecommand{\bibinfo}[2]{#2}
\ifx\xfnm\relax \def\xfnm[#1]{\unskip,\space#1}\fi
\bibitem[{Obied et~al.(2018)Obied, Ooguri, Spodyneiko, and
  Vafa}]{Obied:2018sgi}
\bibinfo{author}{G.~Obied}, \bibinfo{author}{H.~Ooguri},
  \bibinfo{author}{L.~Spodyneiko}, \bibinfo{author}{C.~Vafa},
\newblock \bibinfo{title}{{De Sitter Space and the Swampland}}
  (\bibinfo{year}{2018}). \href{http://arxiv.org/abs/1806.08362}{{\tt
  arXiv:1806.08362}}.
\bibitem[{Garg and Krishnan(2018)}]{Garg:2018reu}
\bibinfo{author}{S.~K. Garg}, \bibinfo{author}{C.~Krishnan},
\newblock \bibinfo{title}{{Bounds on Slow Roll and the de Sitter Swampland}}
  (\bibinfo{year}{2018}). \href{http://arxiv.org/abs/1807.05193}{{\tt
  arXiv:1807.05193}}.
\bibitem[{Ooguri et~al.(2019)Ooguri, Palti, Shiu, and Vafa}]{Ooguri:2018wrx}
\bibinfo{author}{H.~Ooguri}, \bibinfo{author}{E.~Palti},
  \bibinfo{author}{G.~Shiu}, \bibinfo{author}{C.~Vafa},
\newblock \bibinfo{title}{{Distance and de Sitter Conjectures on the
  Swampland}},
\newblock \bibinfo{journal}{Phys. Lett.} \bibinfo{volume}{B788}
  (\bibinfo{year}{2019}) \bibinfo{pages}{180--184}.
  \DOIprefix\doi{10.1016/j.physletb.2018.11.018}.
  \href{http://arxiv.org/abs/1810.05506}{{\tt arXiv:1810.05506}}.
\bibitem[{Kachru and Trivedi(2019)}]{Kachru:2018aqn}
\bibinfo{author}{S.~Kachru}, \bibinfo{author}{S.~P. Trivedi},
\newblock \bibinfo{title}{{A comment on effective field theories of flux
  vacua}},
\newblock \bibinfo{journal}{Fortsch. Phys.} \bibinfo{volume}{67}
  (\bibinfo{year}{2019}) \bibinfo{pages}{1800086}.
  \DOIprefix\doi{10.1002/prop.201800086}.
  \href{http://arxiv.org/abs/1808.08971}{{\tt arXiv:1808.08971}}.
\bibitem[{Akrami et~al.(2019)Akrami, Kallosh, Linde, and
  Vardanyan}]{Akrami:2018ylq}
\bibinfo{author}{Y.~Akrami}, \bibinfo{author}{R.~Kallosh},
  \bibinfo{author}{A.~Linde}, \bibinfo{author}{V.~Vardanyan},
\newblock \bibinfo{title}{{The Landscape, the Swampland and the Era of
  Precision Cosmology}},
\newblock \bibinfo{journal}{Fortsch. Phys.} \bibinfo{volume}{67}
  (\bibinfo{year}{2019}) \bibinfo{pages}{1800075}.
  \DOIprefix\doi{10.1002/prop.201800075}.
  \href{http://arxiv.org/abs/1808.09440}{{\tt arXiv:1808.09440}}.
\bibitem[{Agrawal et~al.(2018)Agrawal, Obied, Steinhardt, and
  Vafa}]{Agrawal:2018own}
\bibinfo{author}{P.~Agrawal}, \bibinfo{author}{G.~Obied},
  \bibinfo{author}{P.~J. Steinhardt}, \bibinfo{author}{C.~Vafa},
\newblock \bibinfo{title}{{On the Cosmological Implications of the String
  Swampland}},
\newblock \bibinfo{journal}{Phys. Lett.} \bibinfo{volume}{B784}
  (\bibinfo{year}{2018}) \bibinfo{pages}{271--276}.
  \DOIprefix\doi{10.1016/j.physletb.2018.07.040}.
  \href{http://arxiv.org/abs/1806.09718}{{\tt arXiv:1806.09718}}.
\bibitem[{Agrawal and Obied(2019)}]{Agrawal:2018rcg}
\bibinfo{author}{P.~Agrawal}, \bibinfo{author}{G.~Obied},
\newblock \bibinfo{title}{{Dark Energy and the Refined de Sitter Conjecture}},
\newblock \bibinfo{journal}{JHEP} \bibinfo{volume}{06} (\bibinfo{year}{2019})
  \bibinfo{pages}{103}. \DOIprefix\doi{10.1007/JHEP06(2019)103}.
  \href{http://arxiv.org/abs/1811.00554}{{\tt arXiv:1811.00554}}.
\bibitem[{Raveri et~al.(2019)Raveri, Hu, and Sethi}]{Raveri:2018ddi}
\bibinfo{author}{M.~Raveri}, \bibinfo{author}{W.~Hu},
  \bibinfo{author}{S.~Sethi},
\newblock \bibinfo{title}{{Swampland Conjectures and Late-Time Cosmology}},
\newblock \bibinfo{journal}{Phys. Rev.} \bibinfo{volume}{D99}
  (\bibinfo{year}{2019}) \bibinfo{pages}{083518}.
  \DOIprefix\doi{10.1103/PhysRevD.99.083518}.
  \href{http://arxiv.org/abs/1812.10448}{{\tt arXiv:1812.10448}}.
\bibitem[{Bedroya and Vafa(2019)}]{Bedroya:2019snp}
\bibinfo{author}{A.~Bedroya}, \bibinfo{author}{C.~Vafa},
\newblock \bibinfo{title}{{Trans-Planckian Censorship and the Swampland}}
  (\bibinfo{year}{2019}). \href{http://arxiv.org/abs/1909.11063}{{\tt
  arXiv:1909.11063}}.
\bibitem[{Martin and Brandenberger(2001)}]{Martin:2000xs}
\bibinfo{author}{J.~Martin}, \bibinfo{author}{R.~H. Brandenberger},
\newblock \bibinfo{title}{{The TransPlanckian problem of inflationary
  cosmology}},
\newblock \bibinfo{journal}{Phys. Rev.} \bibinfo{volume}{D63}
  (\bibinfo{year}{2001}) \bibinfo{pages}{123501}.
  \DOIprefix\doi{10.1103/PhysRevD.63.123501}.
  \href{http://arxiv.org/abs/hep-th/0005209}{{\tt arXiv:hep-th/0005209}}.
\bibitem[{Easther et~al.(2001)Easther, Greene, Kinney, and
  Shiu}]{Easther:2001fi}
\bibinfo{author}{R.~Easther}, \bibinfo{author}{B.~R. Greene},
  \bibinfo{author}{W.~H. Kinney}, \bibinfo{author}{G.~Shiu},
\newblock \bibinfo{title}{{Inflation as a probe of short distance physics}},
\newblock \bibinfo{journal}{Phys. Rev.} \bibinfo{volume}{D64}
  (\bibinfo{year}{2001}) \bibinfo{pages}{103502}.
  \DOIprefix\doi{10.1103/PhysRevD.64.103502}.
  \href{http://arxiv.org/abs/hep-th/0104102}{{\tt arXiv:hep-th/0104102}}.
\bibitem[{Bedroya et~al.(2019)Bedroya, Brandenberger, Loverde, and
  Vafa}]{Bedroya:2019tba}
\bibinfo{author}{A.~Bedroya}, \bibinfo{author}{R.~Brandenberger},
  \bibinfo{author}{M.~Loverde}, \bibinfo{author}{C.~Vafa},
\newblock \bibinfo{title}{{Trans-Planckian Censorship and Inflationary
  Cosmology}}  (\bibinfo{year}{2019}).
  \href{http://arxiv.org/abs/1909.11106}{{\tt arXiv:1909.11106}}.
\bibitem[{Tenkanen(2019)}]{Tenkanen2019}
\bibinfo{author}{T.~Tenkanen},
\newblock \bibinfo{title}{{Trans-Planckian Censorship, Inflation and Dark
  Matter}}  (\bibinfo{year}{2019}). \href{http://arxiv.org/abs/1910.00521}{{\tt
  arXiv:1910.00521}}.
\bibitem[{Cai and Piao(2019)}]{Cai:2019hge}
\bibinfo{author}{Y.~Cai}, \bibinfo{author}{Y.-S. Piao},
\newblock \bibinfo{title}{{Pre-inflation and Trans-Planckian Censorship}}
  (\bibinfo{year}{2019}). \href{http://arxiv.org/abs/1909.12719}{{\tt
  arXiv:1909.12719}}.
\bibitem[{Scalisi and Valenzuela(2019)}]{Scalisi:2018eaz}
\bibinfo{author}{M.~Scalisi}, \bibinfo{author}{I.~Valenzuela},
\newblock \bibinfo{title}{{Swampland distance conjecture, inflation and
  $\alpha$-attractors}},
\newblock \bibinfo{journal}{JHEP} \bibinfo{volume}{08} (\bibinfo{year}{2019})
  \bibinfo{pages}{160}. \DOIprefix\doi{10.1007/JHEP08(2019)160}.
  \href{http://arxiv.org/abs/1812.07558}{{\tt arXiv:1812.07558}}.
\bibitem[{Ashoorioon(2019)}]{Ashoorioon:2018sqb}
\bibinfo{author}{A.~Ashoorioon},
\newblock \bibinfo{title}{{Rescuing Single Field Inflation from the
  Swampland}},
\newblock \bibinfo{journal}{Phys. Lett.} \bibinfo{volume}{B790}
  (\bibinfo{year}{2019}) \bibinfo{pages}{568--573}.
  \DOIprefix\doi{10.1016/j.physletb.2019.02.009}.
  \href{http://arxiv.org/abs/1810.04001}{{\tt arXiv:1810.04001}}.
\bibitem[{Das(2019)}]{Das:2018hqy}
\bibinfo{author}{S.~Das},
\newblock \bibinfo{title}{{Note on single-field inflation and the swampland
  criteria}},
\newblock \bibinfo{journal}{Phys. Rev.} \bibinfo{volume}{D99}
  (\bibinfo{year}{2019}) \bibinfo{pages}{083510}.
  \DOIprefix\doi{10.1103/PhysRevD.99.083510}.
  \href{http://arxiv.org/abs/1809.03962}{{\tt arXiv:1809.03962}}.
\bibitem[{Mizuno et~al.(2019)Mizuno, Mukohyama, Pi, and Zhang}]{Mizuno:2019pcm}
\bibinfo{author}{S.~Mizuno}, \bibinfo{author}{S.~Mukohyama},
  \bibinfo{author}{S.~Pi}, \bibinfo{author}{Y.-L. Zhang},
\newblock \bibinfo{title}{{Hyperbolic field space and swampland conjecture for
  DBI scalar}},
\newblock \bibinfo{journal}{JCAP} \bibinfo{volume}{1909} (\bibinfo{year}{2019})
  \bibinfo{pages}{072}. \DOIprefix\doi{10.1088/1475-7516/2019/09/072}.
  \href{http://arxiv.org/abs/1905.10950}{{\tt arXiv:1905.10950}}.
\bibitem[{Akrami et~al.(2018)}]{Akrami:2018odb}
\bibinfo{author}{Y.~Akrami}, et~al. (\bibinfo{collaboration}{Planck}),
\newblock \bibinfo{title}{{Planck 2018 results. X. Constraints on inflation}}
  (\bibinfo{year}{2018}). \href{http://arxiv.org/abs/1807.06211}{{\tt
  arXiv:1807.06211}}.
\bibitem[{Fukuda et~al.(2019)Fukuda, Saito, Shirai, and
  Yamazaki}]{Fukuda:2018haz}
\bibinfo{author}{H.~Fukuda}, \bibinfo{author}{R.~Saito},
  \bibinfo{author}{S.~Shirai}, \bibinfo{author}{M.~Yamazaki},
\newblock \bibinfo{title}{{Phenomenological Consequences of the Refined
  Swampland Conjecture}},
\newblock \bibinfo{journal}{Phys. Rev.} \bibinfo{volume}{D99}
  (\bibinfo{year}{2019}) \bibinfo{pages}{083520}.
  \DOIprefix\doi{10.1103/PhysRevD.99.083520}.
  \href{http://arxiv.org/abs/1810.06532}{{\tt arXiv:1810.06532}}.
\bibitem[{Das(2019)}]{Das:2018rpg}
\bibinfo{author}{S.~Das},
\newblock \bibinfo{title}{{Warm Inflation in the light of Swampland Criteria}},
\newblock \bibinfo{journal}{Phys. Rev.} \bibinfo{volume}{D99}
  (\bibinfo{year}{2019}) \bibinfo{pages}{063514}.
  \DOIprefix\doi{10.1103/PhysRevD.99.063514}.
  \href{http://arxiv.org/abs/1810.05038}{{\tt arXiv:1810.05038}}.
\bibitem[{Motaharfar et~al.(2019)Motaharfar, Kamali, and
  Ramos}]{Motaharfar:2018zyb}
\bibinfo{author}{M.~Motaharfar}, \bibinfo{author}{V.~Kamali},
  \bibinfo{author}{R.~O. Ramos},
\newblock \bibinfo{title}{{Warm inflation as a way out of the swampland}},
\newblock \bibinfo{journal}{Phys. Rev.} \bibinfo{volume}{D99}
  (\bibinfo{year}{2019}) \bibinfo{pages}{063513}.
  \DOIprefix\doi{10.1103/PhysRevD.99.063513}.
  \href{http://arxiv.org/abs/1810.02816}{{\tt arXiv:1810.02816}}.
\bibitem[{Berera(1995)}]{Berera:1995ie}
\bibinfo{author}{A.~Berera},
\newblock \bibinfo{title}{{Warm inflation}},
\newblock \bibinfo{journal}{Phys. Rev. Lett.} \bibinfo{volume}{75}
  (\bibinfo{year}{1995}) \bibinfo{pages}{3218--3221}.
  \DOIprefix\doi{10.1103/PhysRevLett.75.3218}.
  \href{http://arxiv.org/abs/astro-ph/9509049}{{\tt arXiv:astro-ph/9509049}}.
\bibitem[{Yokoyama and Linde(1999)}]{Yokoyama:1998ju}
\bibinfo{author}{J.~Yokoyama}, \bibinfo{author}{A.~D. Linde},
\newblock \bibinfo{title}{{Is warm inflation possible?}},
\newblock \bibinfo{journal}{Phys. Rev.} \bibinfo{volume}{D60}
  (\bibinfo{year}{1999}) \bibinfo{pages}{083509}.
  \DOIprefix\doi{10.1103/PhysRevD.60.083509}.
  \href{http://arxiv.org/abs/hep-ph/9809409}{{\tt arXiv:hep-ph/9809409}}.
\bibitem[{Bastero-Gil et~al.(2016)Bastero-Gil, Berera, Ramos, and
  Rosa}]{Bastero-Gil:2016qru}
\bibinfo{author}{M.~Bastero-Gil}, \bibinfo{author}{A.~Berera},
  \bibinfo{author}{R.~O. Ramos}, \bibinfo{author}{J.~G. Rosa},
\newblock \bibinfo{title}{{Warm Little Inflaton}},
\newblock \bibinfo{journal}{Phys. Rev. Lett.} \bibinfo{volume}{117}
  (\bibinfo{year}{2016}) \bibinfo{pages}{151301}.
  \DOIprefix\doi{10.1103/PhysRevLett.117.151301}.
  \href{http://arxiv.org/abs/1604.08838}{{\tt arXiv:1604.08838}}.
\bibitem[{Bastero-Gil et~al.(2019)Bastero-Gil, Berera, Ramos, and
  Rosa}]{Bastero-Gil:2019gao}
\bibinfo{author}{M.~Bastero-Gil}, \bibinfo{author}{A.~Berera},
  \bibinfo{author}{R.~O. Ramos}, \bibinfo{author}{J.~G. Rosa},
\newblock \bibinfo{title}{{Towards a reliable effective field theory of
  inflation}}  (\bibinfo{year}{2019}).
  \href{http://arxiv.org/abs/1907.13410}{{\tt arXiv:1907.13410}}.
\bibitem[{Baumann and McAllister(2015)}]{Baumann:2014nda}
\bibinfo{author}{D.~Baumann}, \bibinfo{author}{L.~McAllister},
  \bibinfo{title}{{Inflation and String Theory}}, Cambridge Monographs on
  Mathematical Physics, \bibinfo{publisher}{Cambridge University Press},
  \bibinfo{year}{2015}. \URLprefix
  \url{http://www.cambridge.org/mw/academic/subjects/physics/theoretical-physics-and-mathematical-physics/inflation-and-string-theory?format=HB}.
  \DOIprefix\doi{10.1017/CBO9781316105733}.
  \href{http://arxiv.org/abs/1404.2601}{{\tt arXiv:1404.2601}}.
\bibitem[{Wu et~al.(2016)}]{Wu:2016hul}
\bibinfo{author}{W.~L.~K. Wu}, et~al.,
\newblock \bibinfo{title}{{Initial Performance of BICEP3: A Degree Angular
  Scale 95 GHz Band Polarimeter}},
\newblock \bibinfo{journal}{J. Low. Temp. Phys.} \bibinfo{volume}{184}
  (\bibinfo{year}{2016}) \bibinfo{pages}{765--771}.
  \DOIprefix\doi{10.1007/s10909-015-1403-x}.
  \href{http://arxiv.org/abs/1601.00125}{{\tt arXiv:1601.00125}}.
\bibitem[{Matsumura et~al.(2013)}]{Matsumura:2013aja}
\bibinfo{author}{T.~Matsumura}, et~al.,
\newblock \bibinfo{title}{{Mission design of LiteBIRD}}
  (\bibinfo{year}{2013}). \DOIprefix\doi{10.1007/s10909-013-0996-1}.
  \href{http://arxiv.org/abs/1311.2847}{{\tt arXiv:1311.2847}},
  \bibinfo{note}{[J. Low. Temp. Phys.176,733(2014)]}.
\bibitem[{Ramos and da~Silva(2013)}]{Ramos:2013nsa}
\bibinfo{author}{R.~O. Ramos}, \bibinfo{author}{L.~A. da~Silva},
\newblock \bibinfo{title}{{Power spectrum for inflation models with quantum and
  thermal noises}},
\newblock \bibinfo{journal}{JCAP} \bibinfo{volume}{1303} (\bibinfo{year}{2013})
  \bibinfo{pages}{032}. \DOIprefix\doi{10.1088/1475-7516/2013/03/032}.
  \href{http://arxiv.org/abs/1302.3544}{{\tt arXiv:1302.3544}}.
\bibitem[{Chowdhury et~al.(2019)Chowdhury, Martin, Ringeval, and
  Vennin}]{Chowdhury:2019otk}
\bibinfo{author}{D.~Chowdhury}, \bibinfo{author}{J.~Martin},
  \bibinfo{author}{C.~Ringeval}, \bibinfo{author}{V.~Vennin},
\newblock \bibinfo{title}{{Inflation after Planck: Judgment Day}}
  (\bibinfo{year}{2019}). \href{http://arxiv.org/abs/1902.03951}{{\tt
  arXiv:1902.03951}}.
\bibitem[{Bastero-Gil et~al.(2018)Bastero-Gil, Berera, Brandenberger, Moss,
  Ramos, and Rosa}]{Bastero-Gil:2016mrl}
\bibinfo{author}{M.~Bastero-Gil}, \bibinfo{author}{A.~Berera},
  \bibinfo{author}{R.~Brandenberger}, \bibinfo{author}{I.~G. Moss},
  \bibinfo{author}{R.~O. Ramos}, \bibinfo{author}{J.~G. Rosa},
\newblock \bibinfo{title}{{The role of fluctuation-dissipation dynamics in
  setting initial conditions for inflation}},
\newblock \bibinfo{journal}{JCAP} \bibinfo{volume}{1801} (\bibinfo{year}{2018})
  \bibinfo{pages}{002}. \DOIprefix\doi{10.1088/1475-7516/2018/01/002}.
  \href{http://arxiv.org/abs/1612.04726}{{\tt arXiv:1612.04726}}.

\end{thebibliography}

\end{document}